\documentclass[fleqn,usenatbib]{mnras}
\usepackage{mathptmx}
\usepackage[varvw]{newtxmath}  %evs: Added to prevent cursive v to look like $\nu$.
\usepackage[T1]{fontenc}
\usepackage{ae,aecompl}
\usepackage{times}
\usepackage{amsmath}
\usepackage{upgreek}
\usepackage{xcolor}
\usepackage{graphicx}
\usepackage{pdflscape}
\usepackage{multicol} 
\usepackage{amsmath,bm}
\usepackage{caption}
\begin{document}

\title{Origin of open clusters revealed by the evolution of the $m_{\rm max} -M_{\rm ecl}$ relation}

\author[J. W. Zhou]{
J. W. Zhou \thanks{E-mail: jwzhou@mpifr-bonn.mpg.de}$^{1}$
Sami Dib $^{2}$
Pavel Kroupa $^{3,4}$
\\
$^{1}$Max-Planck-Institut f\"{u}r Radioastronomie, Auf dem H\"{u}gel 69, 53121 Bonn, Germany\\
% $^{2}$Max Planck Institute f\"{u}r Astronomie, K\"{o}nigstuhl 17, 69117 Heidelberg, Germany\\
$^{2}$Max-Panck-Institut f\"{u}r Astronomie, K\"{o}nigstuhl 17, 69117, Heidelberg, Germany\\
$^{3}$Helmholtz-Institut f{\"u}r Strahlen- und Kernphysik (HISKP), Universität Bonn, Nussallee 14–16, 53115 Bonn, Germany\\
$^{4}$Charles University in Prague, Faculty of Mathematics and Physics, Astronomical Institute, V Hole{\v s}ovi{\v c}k{\'a}ch 2, CZ-180 00 Praha 8, Czech Republic}

\date{Accepted XXX. Received YYY; in original form ZZZ}

\pubyear{2024}
% Don't change these lines
% \label{firstpage}
% \pagerange{\pageref{firstpage}--\pageref{lastpage}}
\maketitle

\begin{abstract}
Using the Gaia DR3 open cluster catalog, we identified the most massive star in each observed cluster. Examining the $m_{\rm max}$–$M_{\rm cluster}$ relations across different age ranges, we find that as clusters age, the relation gradually deviates from the initial $m_{\rm max}$–$M_{\rm ecl}$ relation and eventually exhibits clear age stratification. We conducted N-body simulations for both individual cluster evolution and subcluster coalescence. Four gas expulsion modes were tested for individual clusters, and two scenarios were modeled for cluster coalescence. Under all four gas expulsion modes, the evolution of the $m_{\rm max}$–$M_{\rm cluster}$ relation follows a similar trajectory, differing mainly in evolutionary speed. The coalescence simulations show comparable behavior but align better with the observations, as both exhibit systematically lower $m_{\rm max}$–$M_{\rm cluster}$ relations than individual cluster simulations. This systematically lower observed $m_{\rm max}$–$M_{\rm cluster}$ relation suggests slower cluster mass loss and smaller masses for the most massive stars—both conditions reproduced in the coalescence simulations. Observations also show that clusters older than 5 Myr have most massive stars significantly deviating from the initial $m_{\rm max}$–$M_{\rm ecl}$ relation. From this perspective, the coalescence simulations also provide a better match to the observations. In conclusion, the evolution of the $m_{\rm max}$–$M_{\rm ecl}$ relation supports subcluster coalescence as a dominant pathway for open cluster formation, consistent with our previous work.

\end{abstract}


% \titlerunning{Isolated evolution from clump to embedded cluster}
% \authorrunning{J. W. Zhou, Pavel Kroupa}

% \maketitle 

\section{Introduction}

The question of whether newborn stars in star clusters are drawn randomly from the initial mass function (IMF) is critical for various fields of astrophysics. The stochastic (or random) sampling approach posits that each star's formation in an embedded cluster is an independent event, with its mass being randomly selected from a probability density function (PDF), effectively treating the IMF as a PDF. Conversely, the optimal sampling theory \citep{Kroupa2013-115, Schulz2015-582} suggests a deterministic relationship between the mass of an embedded cluster and the mass of each individual star within it, due to a self-regulation star formation process in the cluster \citep{Kroupa2015-93,Yan2023-670}.
A key factor in determining whether a star cluster should be modeled using stochastic sampling or optimal sampling from the IMF is the presence of a significant relationship between the mass of the most massive star ($m_{\rm max}$) and the total mass of the embedded cluster in stars ($M_{\rm ecl}$). If a statistically significant correlation exists between $m_{\rm max}$ and $M_{\rm ecl}$, it would suggest that physical processes, such as self-regulation during star formation within the cluster, are at play. Actually,
the $m_{\rm max} -M_{\rm ecl}$ relation has been well established analytically \citep{Elmegreen2000-539,Weidner2004-348,Oey2005-620}, observationally \citep{Larson2003-287,Weidner2006-365, Weidner2010-401,Zhou2024PASP-2} and numerically \citep{Bonnell2003-343, Bonnell2004-349,Peters2011-729}.
\citet{Yan2023-670} provided a deterministic tool for sampling stellar mass that reproduces the observed $m_{\rm max} -M_{\rm ecl}$ relation and aligns well with the luminosities of star-forming molecular clumps (also refer to \citet{Zhou2024PASP-1}). 
The embedded star cluster samples in \citet{Yan2023-670} are younger than 5 Myr. In this study, we explore how the initial $m_{\rm max} - M_{\rm ecl}$ relation for embedded clusters evolves over time, with a particular focus on the similarities and differences between this relation and the $m_{\rm max} - M_{\rm cluster}$ relation for open clusters. Furthermore, we discuss how the temporal variations in the initial $m_{\rm max} - M_{\rm ecl}$ relation reflect the dynamics of cluster evolution and the formation processes of open clusters.

\section{Sample and Simulation data}

\subsection{Sample}

\citet{Hunt2023-673}
conducted a blind, all-sky search for open clusters using 729 million sources from Gaia DR3 down to magnitude G$\approx$20, creating a homogeneous catalog of 7167 clusters. 
By measuring cluster masses and Jacobi radii, the members in the star cluster catalog are classified into two categories, i,e, bound open clusters (OCs) and unbound moving groups (MGs) \citep{Hunt2024-686}.
In \citet{Hunt2024-686},
the catalog includes 5647 OCs and 1309 MGs, where 3530 OCs and 539 MGs are of high-quality with median color-magnitude diagram (CMD) classifications greater than 0.5 and signal-to-noise ratio (S/N) greater than 5$\sigma$ \citep{Hunt2023-673}. 
% For the 3530 high-quality OCs, most of them (3103) have the distance less than 3.6~kpc. 
In this work, we only focus on high-quality OCs and MGs.
To calculate the photometric masses of the clusters, \citet{Hunt2024-686} first derived the photometric masses of the member stars in each cluster using the PARSEC isochrones \citep{Bressan2012-427}. They then corrected for selection effects and applied a correction for unresolved binary stars. Subsequently, mass functions were fitted and integrated to determine the total cluster mass.

\begin{table}
\centering
% \tablenum{1}
\caption{N-body simulations modeling both the evolution of individual clusters and the coalescence of multiple clusters under different gas expulsion modes. More details can refer to \citet{Zhou2024-691-293} and \citet{Zhou2025-537-845}.
"Time" is the time when the simulation is terminated. The cluster may or may not have dissolved by this time. 
% We consider two types of gas expulsion modes, i.e. "fast" and "moderate".
% i.e. "fast" ($\tau_g$) and "moderate" (5$\tau_g$). $\tau_g$ is the gas depletion time of the gas expulsion process. 
}
\label{tab1}
\begin{tabular}{cccc}
\hline
Cases	&	Mass (M$_{\odot}$)	&	Time (Myr)	&	Dissolved	\\
10000-fast	&	10000	&	263	&	yes	\\
10000-moderate1	&	10000	&	100	&	no	\\
10000-moderate2	&	10000	&	270	&	no	\\
10000-slow	&	10000	&	359.25	&	no	\\
3000-fast	&	3000	&	100	&	yes	\\
3000-moderate1	&	3000	&	441	&	no	\\
3000-moderate2	&	3000	&	256.25	&	yes	\\
3000-slow	&	3000	&	950.25	&	yes	\\
1000-fast	&	1000	&	67.5	&	yes	\\
1000-moderate1	&	1000	&	145.5	&	yes	\\
1000-moderate2	&	1000	&	78	&	yes	\\
1000-slow	&	1000	&	533.5	&	yes	\\
300-fast	&	300	&	69	&	yes	\\
300-moderate1	&	300	&	100	&	yes	\\
300-moderate2	&	300	&	45.25	&	yes	\\
300-slow	&	300	&	346.5	&	yes	\\
NGC1893-fast	&	468	&	305.75	&	yes	\\
NGC1893-fast-vd	&	468	&	283.25	&	yes	\\
NGC6334-fast	&	2207	&	614.25	&	yes	\\
NGC6334-fast-vd	&	2207	&	463.75	&	yes	\\
Carina-fast	&	6271	&	556.5	&	no	\\
Carina-fast-vd	&	6271	&	324.5	&	yes	\\
\hline
\label{tab1}
\end{tabular}
\end{table}

% \begin{figure}
% \centering
% \includegraphics[width=0.48\textwidth]{fig/slow.pdf}
% \caption{Fitting the observed $m_{\rm max} -M_{\rm cluster}$ relations of star clusters using N-body simulations under slow gas expulsion mode.}
% \label{slow}
% \end{figure}
\subsection{N-body simulations}

A detailed description of the simulation methodology is provided in Appendix.\ref{nbody}.
% For the convenience of the reader, these setups are summarized again in Appendix.\ref{nbody}.
In \citet{Zhou2024-691-293},
we computed the dynamical evolution of three clusters with the masses in stars of 300 M$_{\odot}$, 1000 M$_{\odot}$ and 3000 M$_{\odot}$ for the first 100 Myr under four different gas expulsion modes. In this work, we continue the simulations from the previous work until the clusters dissolve, as shown in Table.\ref{tab1}. We also simulate a 10000 M$_{\odot}$ cluster using the same recipe as in \citet{Zhou2024-691-293}.
In \citet{Zhou2025-537-845}, we simulated the post-gas expulsion coalescence of embedded clusters formed within the same parental molecular cloud that undergo a phase of expansion caused by the effects of gas expulsion. 
% In this work, we only consider two extreme cases in the coalescence simulation, i.e. "fast" and "fast-vd" \footnote{\textcolor{magenta}{As presented in \citet{Zhou2025-537-845} and Appendix.\ref{nbody}, in the 'fast' case, all embedded clusters lie in the same $XY$ plane and are stationary relative to one another. In contrast, in the 'fast-vd' case, the embedded clusters are spatially separated along the $Z$-axis; they are distributed in three-dimensional space and possess relative velocities.}}. The simulation results from other parameter choices should lie in between these two extremes, as discussed in \citet{Zhou2025-537-845}.
% Our templates for the initial conditions are the observed spatial, kinematic, and mass distributions of subclusters in three representative massive star-forming regions (MSFRs), namely NGC 1893, NGC 6334, and the Carina Nebula. 

\begin{figure}
\centering
\includegraphics[width=0.45\textwidth]{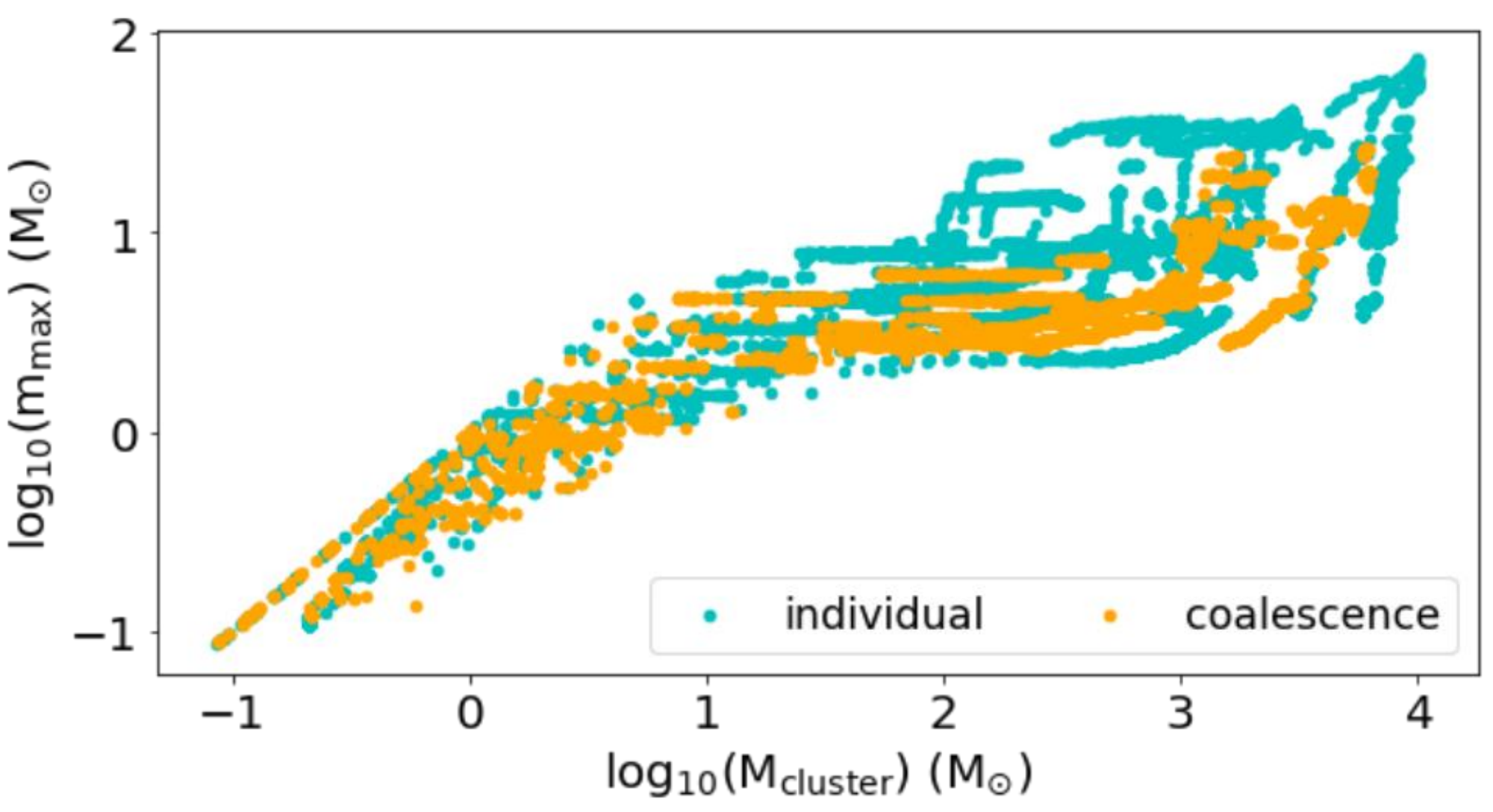}
\caption{Comparison of the $m_{\rm max} -M_{\rm cluster}$ relations in the cluster coalescence and individual cluster simulations shown in Table.\ref{tab1}. The timestep is linearly spaced at 0.25 Myr.}
\label{Mmt} 
\end{figure}

Overall, the simulations can be divided into two categories: the cluster coalescence simulations \citep{Zhou2025-537-845} and the individual cluster simulations \citep{Zhou2024-691-293}. 
% i.e. "individual" and "coalescence" in Table.\ref{all}. 
As described in Appendix.\ref{nbody}, for individual clusters, we simulated four gas expulsion modes (fast, moderate2/mod2, moderate1/mod1, and slow), characterized by gas expulsion timescales of $\tau_g$, 2$\tau_g$, 5$\tau_g$, and 10$\tau_g$, respectively (see Appendix.\ref{nbody} for the definition of $\tau_g$). To mitigate the stochasticity of the simulation results, we further refined the mass grid and simulated additional clusters using the same setups. For the clusters in Table.\ref{tab1}, as shown in Fig.\ref{Mmt}, the $m_{\rm max} -M_{\rm cluster}$ relations of the cluster coalescence and individual cluster simulations show a clear difference at the beginning, but the discrepancy decreases later. Therefore, all new simulations  were evolved until 100 Myr. 
%in order to reduce computational costs, (i.e., except those listed in Table.\ref{all})
Finally, the simulated individual cluster masses are [300, 562, 1000, 1333, 1778, 2371, 3000, 5623, 10000] M$_{\odot}$, corresponding to [10$^{2.5}$, 10$^{2.75}$, 10$^{3}$, 10$^{3.125}$, 10$^{3.25}$, 10$^{3.375}$, 10$^{3.5}$, 10$^{3.75}$, 10$^{4}$] M$_{\odot}$. 
In the coalescence simulations, each subcluster adopts the fast gas expulsion mode. As described in Appendix.\ref{nbody}, we considered two scenarios: (i) all subclusters are initially at rest and located in the same plane ("-fast"), and (ii) the subclusters are spatially separated and possess velocity differences ("-fast-vd"). In this owrk, we  further simulated other five cluster complexes to 100 Myr identified in \citet{Zhou2024-688L}, their masses are 2628 (M17), 2118 (NGC6357), 1015 (Eagle), 658 (Lagoon) and 314 (NGC2264) M$_{\odot}$, respectively.

\section{Results and Discussion}

\subsection{The $m_{\rm max} -M_{\rm cluster}$ relation in the observations}

\begin{figure*}
\centering
\includegraphics[width=0.9\textwidth]{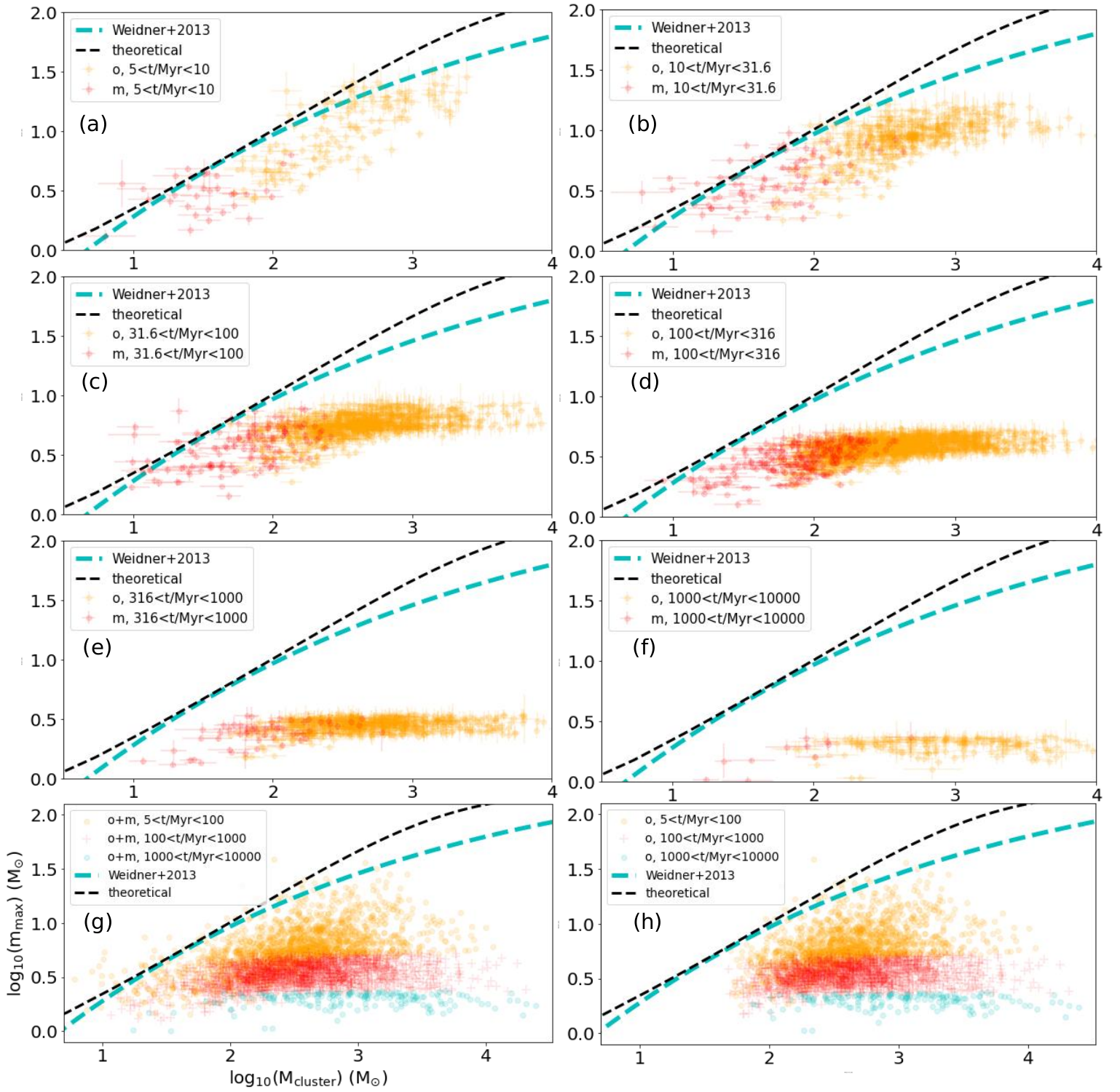}
\caption{
The $m_{\rm max} -M_{\rm cluster}$ relation of the observed open clusters evolves over time. The observational and theoretical $m_{\rm max} -M_{\rm ecl}$ relations in equation.\ref{eq:mmaxmecl2} and \citet{Zhou2024PASP-2} correspond to the cyan and black dashed curves, respectively. "o" and "m" are the bound open clusters and unbound moving groups in the catalog of \citet{Hunt2024-686}.}
\label{mMo} 
\end{figure*}

We picked out the most massive star in each observed open cluster. In this work, we considered clusters with a 50th percentile age greater than 5 Myr, and required that the mass uncertainty of the most massive star is less than 50\% of its mass.
Fig.\ref{mMo} displays the $m_{\rm max} -M_{\rm cluster}$ relations of open clusters in different age ranges. Overall, as the cluster ages increase, the $m_{\rm max} -M_{\rm cluster}$ relation gradually deviates from the initial $m_{\rm max} -M_{\rm ecl}$ relation, and ultimately manifests as age stratification.
The clusters are gradually losing the most massive stars through evolution. 
The median distance of all clusters is $\approx$ 2 kpc. In Fig.\ref{dis}, we divided them into two groups and found that their $m_{\rm max} -M_{\rm ecl}$ relations are mixed together, indicating that the distance factor does not affect the analysis in this work.
% Those massive stars may have been ejected from the cluster due to dynamical interactions, or they may have evolved into supernovae or later dense objects, or merger of clusters. 

% The dataset in \citet{Weidner2013-434}, used to fit the initial $m_{\rm max} -M_{\rm ecl}$ relation, requires clusters with age estimates of less than 5 Myr. However, some of these estimates come with significant uncertainties, suggesting that certain clusters could be as old as 9 Myr within the margin of error \citep{Yan2023-670}. In Fig.\ref{mMo}(a), we focus on the open clusters younger than 5 Myr. The $m_{\rm max} -M_{\rm cluster}$ relation of these clusters roughly follows the initial $m_{\rm max} -M_{\rm ecl}$ relation. 

\subsection{The $m_{\rm max} -M_{\rm cluster}$ relation in the simulations}\label{mMsimu}

\begin{table}
\centering
% \tablenum{1}
\caption{The cluster complexes are identified in \citet{Zhou2024-688L}, see also \citet{Kuhn2014-787}.
$M_{\rm tot}$ and $M_{\rm max}$ are the total mass of the cluster complex and the mass of the most massive subcluster in the cluster complex. $m_{\rm max,tot}$ and $m_{\rm max,max}$ are the masses of the most massive stars calculated by the $m_{\rm max} -M_{\rm ecl}$ relation in \citet{Weidner2013-434} using $M_{\rm tot}$ and $M_{\rm max}$, respectively. $m_{\rm max,max}$ is actually used in the coalescence simulations.}
\begin{tabular}{ccccc}
\hline
Complexes	&	$M_{\rm tot}$ 	&	$M_{\rm max}$ &	$m_{\rm max,tot}$ 	&	$m_{\rm max,max}$ 	\\
&($M_{\odot}$) &($M_{\odot}$) &($M_{\odot}$) & ($M_{\odot}$)\\
Carina	&	6271	&	801	&	54&	 26\\
NGC6334 &	2207	&	673	&	38&	 24\\
NGC1893 &	468		&	106	&	20&	 9\\
M17     &	2628	&	1188&	41&	 30\\
NGC6357 &	2118	&	854	&	38&	 26\\
Eagle   &	1015	&	163	&	28&	 12\\
Lagoon  &	658	    &	190	&	24&	 13\\
NGC2264 &	314	    &	75	&	17&	 7\\
\hline
\label{tab2}
\end{tabular}
\end{table}

The evolution of clusters under different simulation scenarios has been extensively analyzed in \citet{Zhou2024-691-293,Zhou2025-537-845,BH}. Here we summarize several key conclusions.
For individual cluster simulations, the extent of cluster mass loss is strongly governed by the gas expulsion mode: the faster the gas expulsion, the greater the mass loss. However, there exists a degeneracy between cluster mass and gas expulsion timescale. 
For example, a massive cluster undergoing fast gas expulsion expands rapidly and loses mass at a high rate, whereas a lower-mass cluster experiencing slow gas expulsion evolves more gradually, resulting in slower mass loss. Consequently, clusters with different initial masses and gas expulsion timescales may display similar evolutionary trajectories on the $M_{\rm cluster}$–t diagram (Fig.\ref{Mti}). In cluster coalescence simulations, the spatial distribution, relative velocities, mass distribution, and gas expulsion modes of subclusters jointly influence both the dynamics of the merging process and the stability of the final product. As shown in Fig.\ref{Mtic}, although each subcluster undergoes fast gas expulsion, the evolution of the coalesced cluster closely resembles that of a single cluster with slow/moderate gas expulsion in the individual cluster simulations. These findings indicate a degeneracy among cluster mass, gas expulsion mode, and the cluster formation channel (i.e., monolithic versus coalescence). 

\begin{figure*}
\centering
\includegraphics[width=1\textwidth]{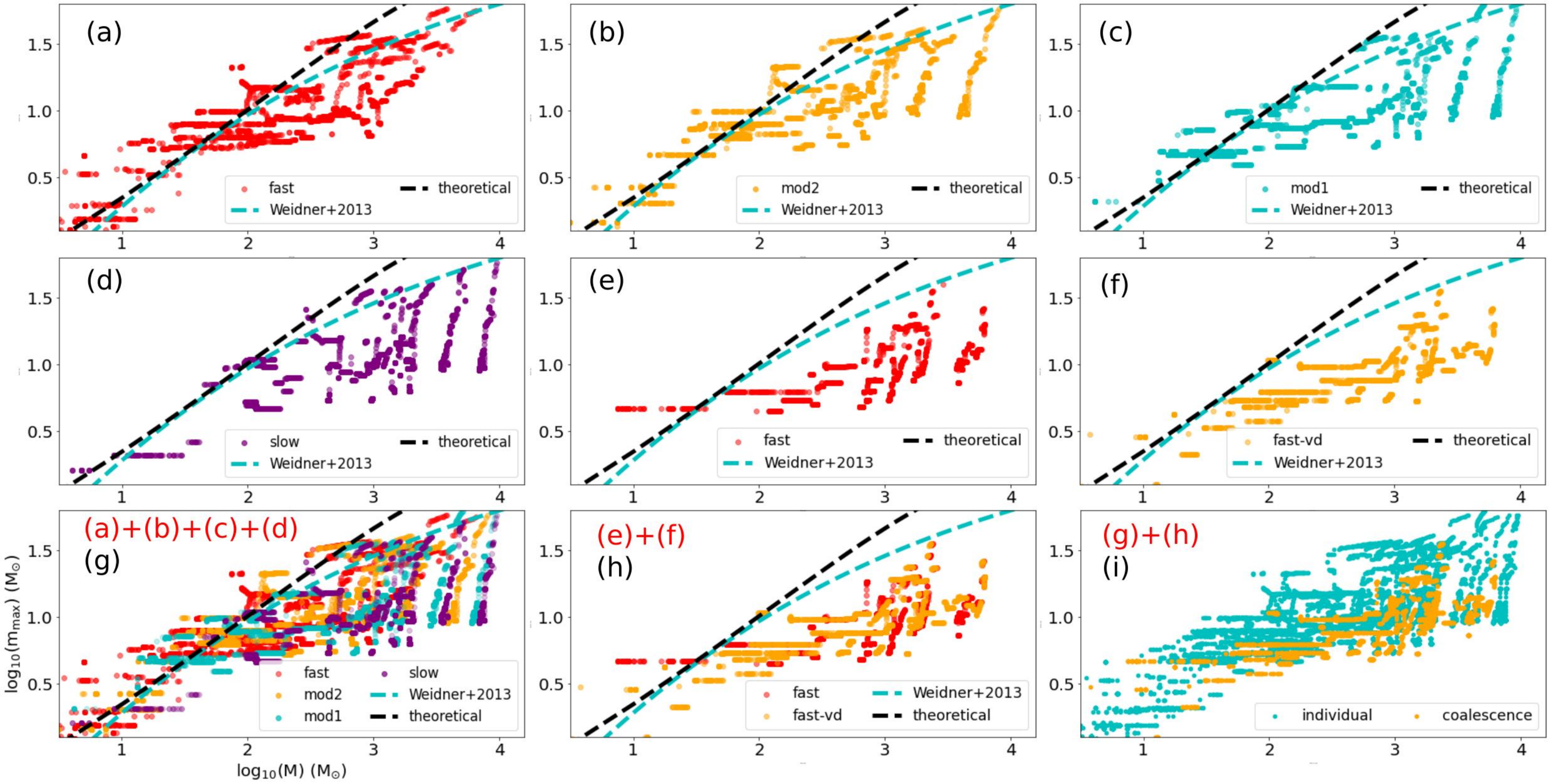}
\caption{Comparison of the $m_{\rm max} -M_{\rm cluster}$ relations in the cluster coalescence and individual cluster simulations within 5-100 Myr. Panels (a)-(d) show the $m_{\rm max} -M_{\rm cluster}$ relations for four different gas expulsion modes in the individual cluster simulations, which are combined in panel (g). Panels (e) and (f) show the $m_{\rm max} -M_{\rm cluster}$ for two scenarios in cluster coalescence simulations, which are merged in panel (h). Panels (g) and (h) are combined in panel (i).
The timestep is linearly spaced at 0.25 Myr.
}
\label{each} 
\end{figure*}

Overall, there are four gas expulsion modes in the individual cluster simulations and two scenarios in the cluster coalescence simulations. As shown in Fig.\ref{each}, under the four different gas expulsion modes, the evolution of the $m_{\rm max} -M_{\rm cluster}$ relation follows a similar trajectory, differing mainly in the speed of evolution. The coalescence simulations show a similar behavior.

\subsection{Comparability of observations and simulations}

\begin{figure}
\centering
\includegraphics[width=0.475\textwidth]{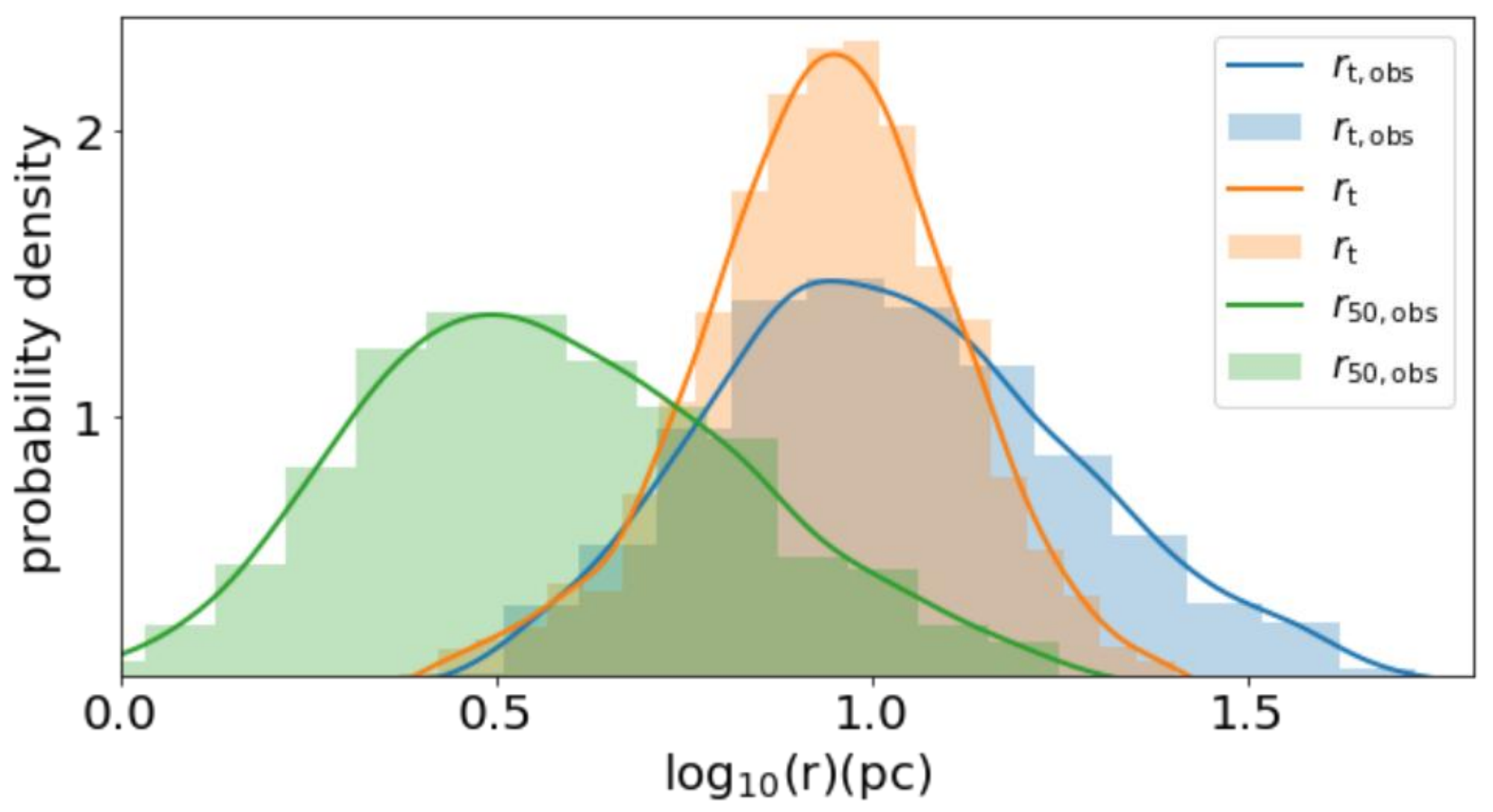}
\caption{Comparison between the observed tidal radius (\( r_{\rm t,obs} \)) and \( r_{\rm 50,obs} \) (the radius containing 50\% of cluster members) from the catalog of \citet{Hunt2024-686}, and the theoretical tidal radius calculated using equation.\ref{tide} with the cluster masses from the same catalog.}
\label{rt} 
\end{figure}

\begin{figure}
\centering
\includegraphics[width=0.475\textwidth]{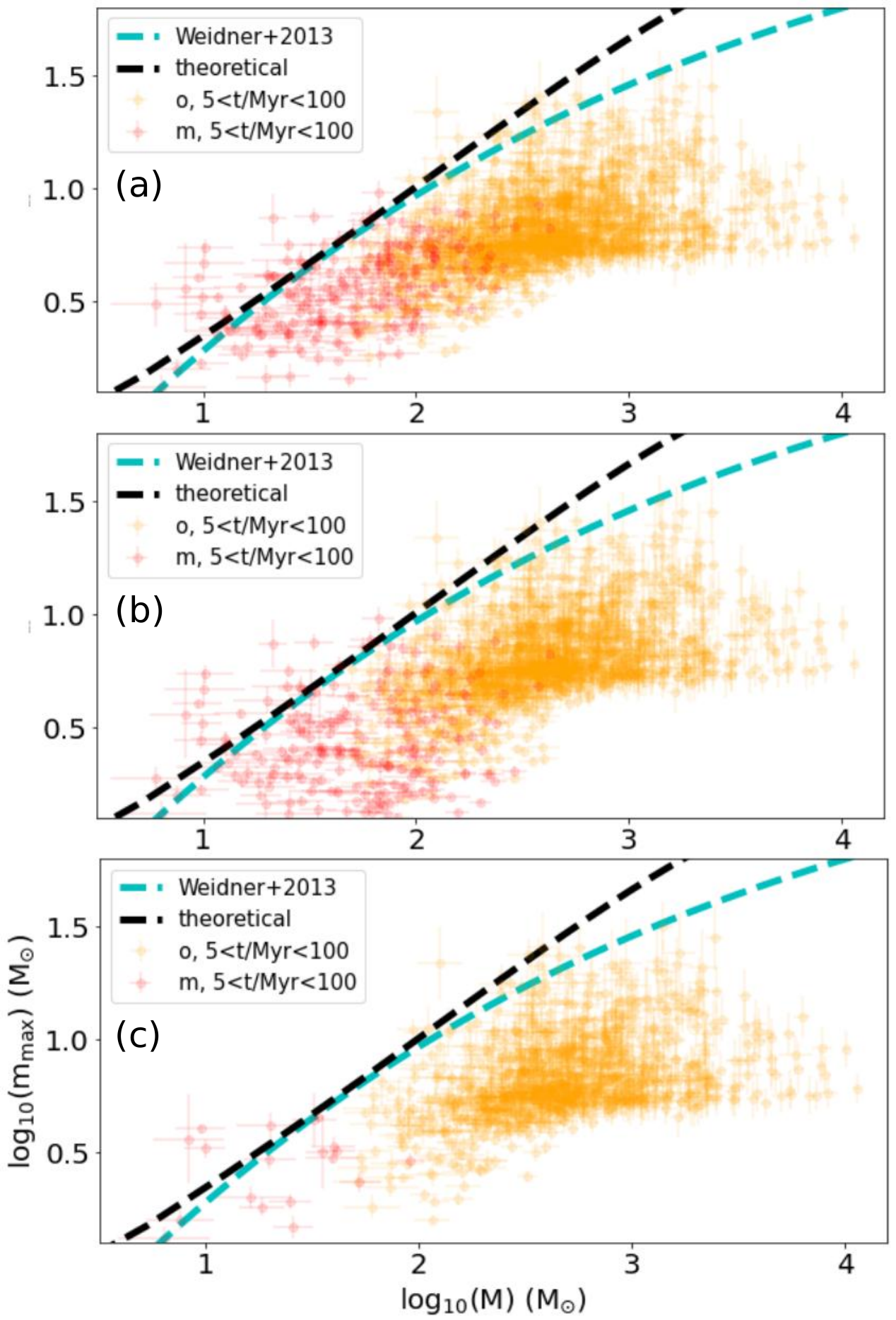}
\caption{The $m_{\rm max} -M_{\rm cluster}$ relation of the observed open clusters within 5-100 Myr.
(a) Same as Fig.\ref{mMo}; (b) Restricting the members of the observed cluster to those within the theoretical tidal radius $r_{\rm t}$, and then selecting the most massive star; (c) Same as panel (b), but further requiring $r_{\rm t}/r_{\rm 50,obs} > 1.69$.}
\label{obs} 
\end{figure}

% \section*{1. Point-mass / Spherically symmetric Galactic potential (King / Roche approximation)}
The simulated clusters travel along circular orbits within the Galaxy, positioned at a galactocentric distance of 8.5 kpc, and are moving at a speed of 220 km s$^{-1}$. The tidal radius \citep{King1962-67} is calculated by
\begin{equation}
r_t \simeq R \left( \frac{M_{\rm cl}}{3\,M_{\rm gal}(<R)} \right)^{1/3}
\label{tide}
\end{equation}
In this equation, the theoretical tidal radius of the cluster, $r_{\rm t}$, is determined by the cluster’s distance from the Galactic center \( R \), its mass $M_{\rm cl}$, and the enclosed Galactic mass \( M_{\rm gal}(<R) \). In each snapshot, we only consider stars with the tidal radius, then calculate the total cluster mass and select the most massive star. 
% This theoretical tidal radius is different from 
The observed tidal radius ($r_{\rm t,obs}$) in the catalog of \citet{Hunt2024-686} is measured as the radius at which the cluster’s stellar density most strongly contrasts with (or begins to exceed) the surrounding field star density. To keep the methodology consistent with that used for calculating physical quantities in the simulated cluster,
based on equation.\ref{tide}, we calculated the theoretical tidal radius for each cluster using the cluster mass in the catalog of \citet{Hunt2024-686}. 
As shown in Fig.\ref{rt}, 
$r_{\rm t,obs}$ is slightly larger than $r_{\rm t}$, and $r_{\rm t}$ is significantly larger than $r_{\rm 50,obs}$ (radius containing 50\% of cluster members). The ratios of the peak values ($r_{\rm t,p}$/$r_{\rm t,obs,p}$ and $r_{\rm t,p}$/$r_{\rm 50,obs,p}$) are 0.93 and 1.69, respectively.
Therefore, the radius $r_{\rm t}$ approximately encloses the total mass of the observed cluster, so we continue to adopt the cluster’s total mass from the catalog. However, we restrict the members of the observed cluster to those within $r_{\rm t}$, and then select the most massive star, in order to maintain consistency with the treatment of the simulated clusters.

Using the new rules, we reexamine the observed $m_{\rm max} -M_{\rm cluster}$ relation within 5-100 Myr presented in Fig.\ref{mMo}.
As a stricter comparison, in another case, we only consider clusters with $r_{\rm t}/r_{\rm 50,obs} > 1.69$. As shown in Fig.\ref{obs}, the observed $m_{\rm max} -M_{\rm cluster}$ relations do not change significantly in all cases.

\subsection{Comparison of observations and simulations}\label{compare}

\begin{figure}
\centering
\includegraphics[width=0.475\textwidth]{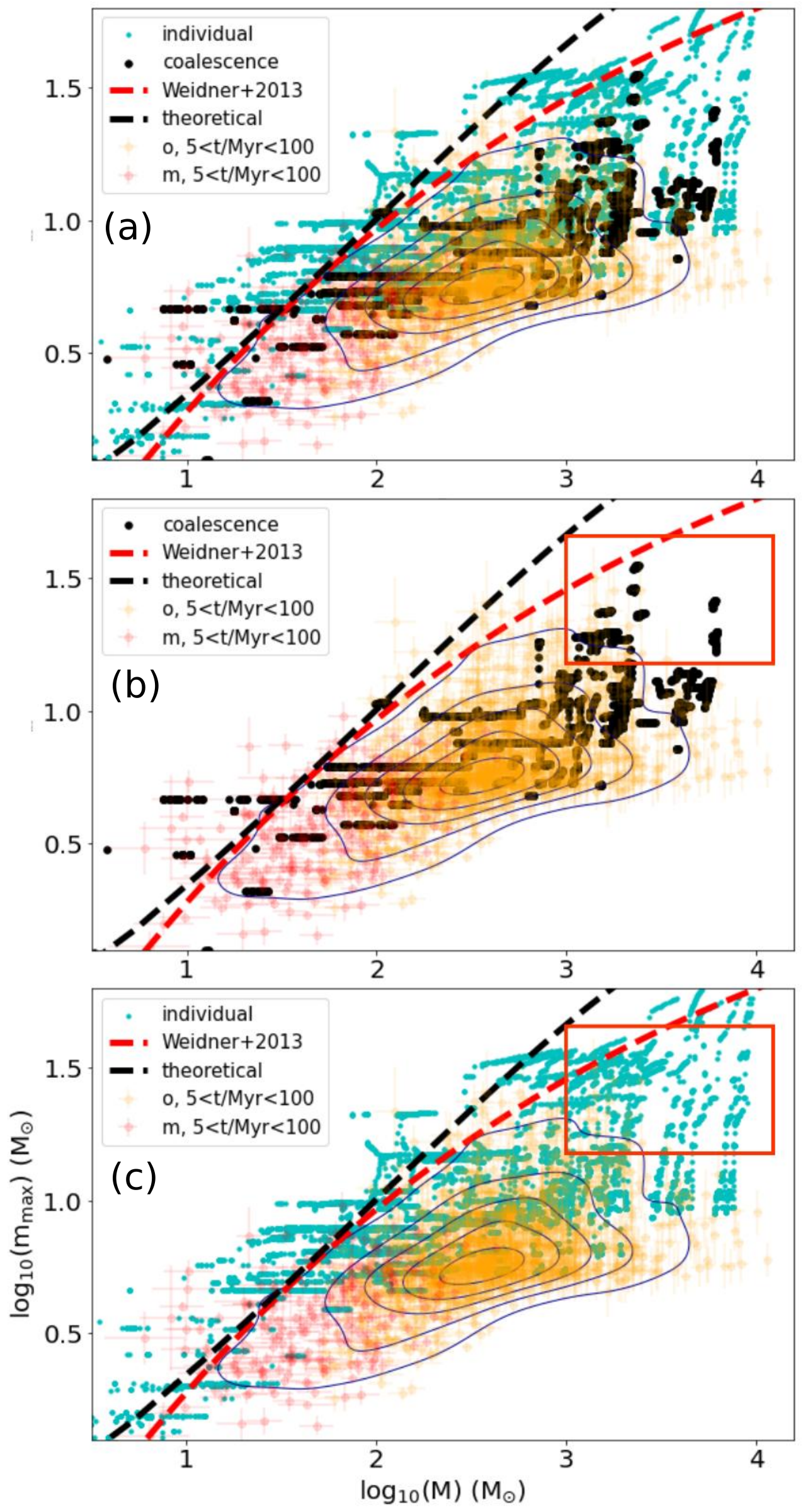}
\caption{Comparison of the $m_{\rm max} -M_{\rm cluster}$ relations between observations and simulations within 5-100 Myr. The blue contours represent the density levels [0.1,0.3,0.5,0.7,0.9] in the kernel density estimation (KDE) plot of the observed $m_{\rm max} -M_{\rm cluster}$ relation.  The observational and theoretical $m_{\rm max} -M_{\rm ecl}$ relations in equation.\ref{eq:mmaxmecl2} and \citet{Zhou2024PASP-2} correspond to the red and black dashed curves, respectively. The red box highlights a region with few observed points.
The timestep is linearly spaced at 0.25 Myr.
}
\label{fit} 
\end{figure}

% The m–M relationship shows a clear difference at the beginning, but the discrepancy decreases later, so the new simulation only focuses on the first 100 Myr. In the following sections, for all simulations, we only consider the first 100 Myr.

In Fig.\ref{fit}, 
the systematically lower $m_{\rm max} -M_{\rm cluster}$ relation in the cluster coalescence simulations is partly attributable to the simulation setup.
The initial $m_{\rm max} -M_{\rm ecl}$ relation of \citet{Weidner2013-434} was applied to each subcluster of the cluster complex. As a result, the most massive star in the cluster complex is notably less massive than the most massive star in an individual cluster with the same total mass, as presented in Table.\ref{tab2}. 
As discussed in Sec.\ref{mMsimu}, the evolution of the coalesced cluster closely resembles that of a single cluster with slow/moderate gas expulsion, which means a slow/moderate mass loss of the cluster. In Fig.\ref{fit}, 
the coalescence simulations agree better with the observations, as both are systematically lower than the individual cluster simulations. The systematically lower observed $m_{\rm max} -M_{\rm cluster}$ relation implies that the clusters lose mass more slowly, and that the most massive stars have smaller masses. Both conditions are satisfied in the coalescence simulations, which therefore agree better with the observations.
% Moreover, the dynamical environments of merging clusters and individual clusters differ substantially. The evolution of the most massive star over time depends not only on the initial content of massive stars, but also on the cluster dynamics.
Furthermore, observations indicate that for clusters with age >5 Myr, the most massive stars have already deviated substantially from the initial $m_{\rm max} -M_{\rm ecl}$ relation. The region marked by the red box in Fig.\ref{fit} only have few observed points. 
However, in the individual cluster simulations, many stars remain in this region. From this perspective, the coalescence simulations also provide a better match to the observations than the individual cluster simulations.
In short, the evolution of the $m_{\rm max} -M_{\rm ecl}$ relation indicates that the subcluster coalescence may be a dominant pathway of open cluster formation. This conclusion is consistent with the results presented in \citet{Zhou2024-691-293,Zhou2025-537-845}, where the coalescence simulations can account for the observed parameter space (e.g., mass, radius and age) of open clusters in the Milky Way.

% However, regarding the evolution of the most massive stars, there are significant differences between the merger and single-cluster simulations. First, cluster complexes lack massive stars; 

% \begin{acknowledgements}
% \end{acknowledgements}
\section*{Acknowledgements}
Thanks to the referee for the detailed and constructive comments, which have significantly contributed to improving this work.

\section{Data availability}
All data used in this work are available from the first author upon request.

\bibliography{ref}
\bibliographystyle{aasjournal}

\begin{appendix}
\twocolumn

\section{Supplementary figures}

\begin{figure}
\centering
\includegraphics[width=0.48\textwidth]{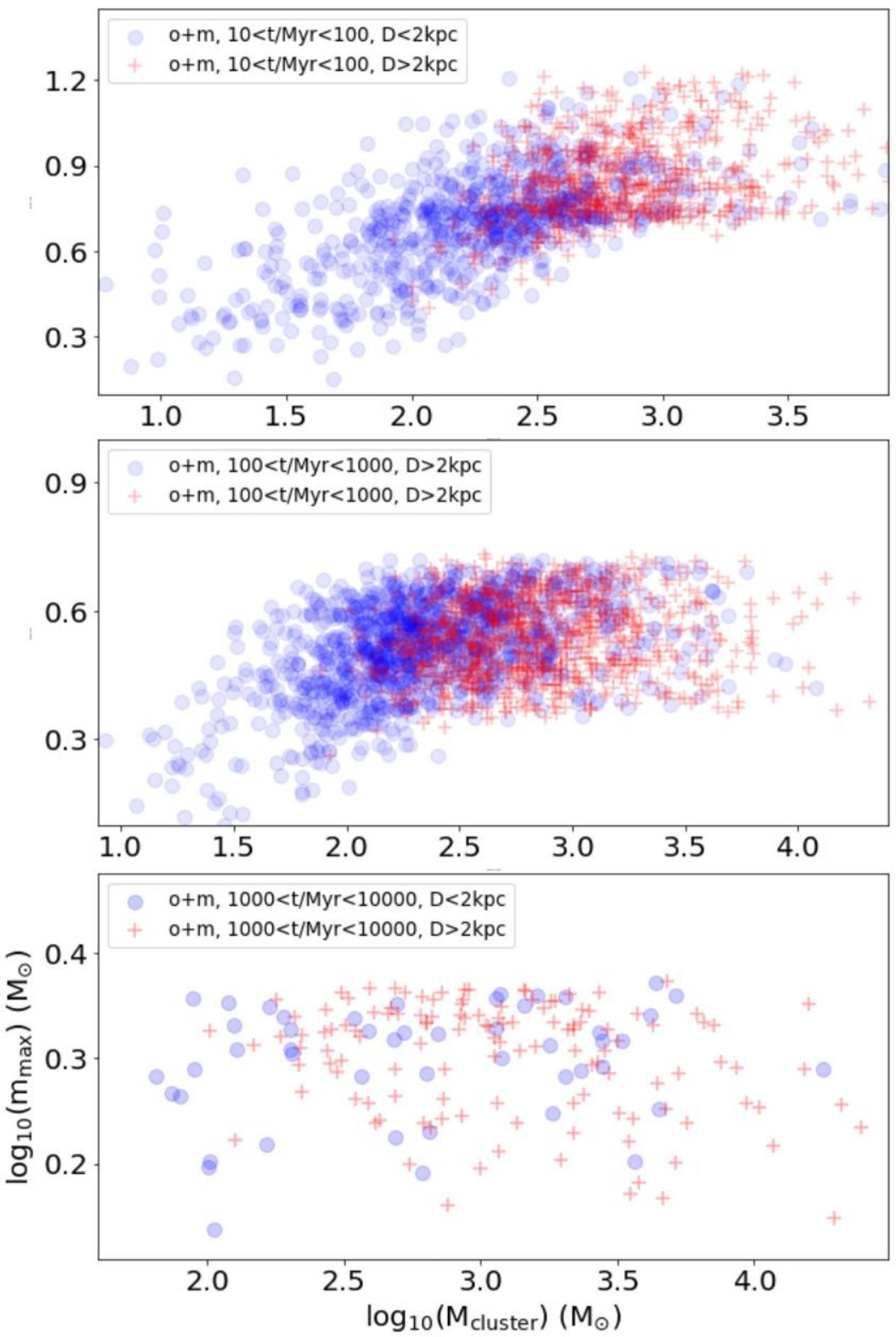}
\caption{
Same as Fig.\ref{mMo}.
The $m_{\rm max} -M_{\rm ecl}$ relations at different ages and distances.}
\label{dis}
\end{figure}

\begin{figure}
\centering
\includegraphics[width=0.475\textwidth]{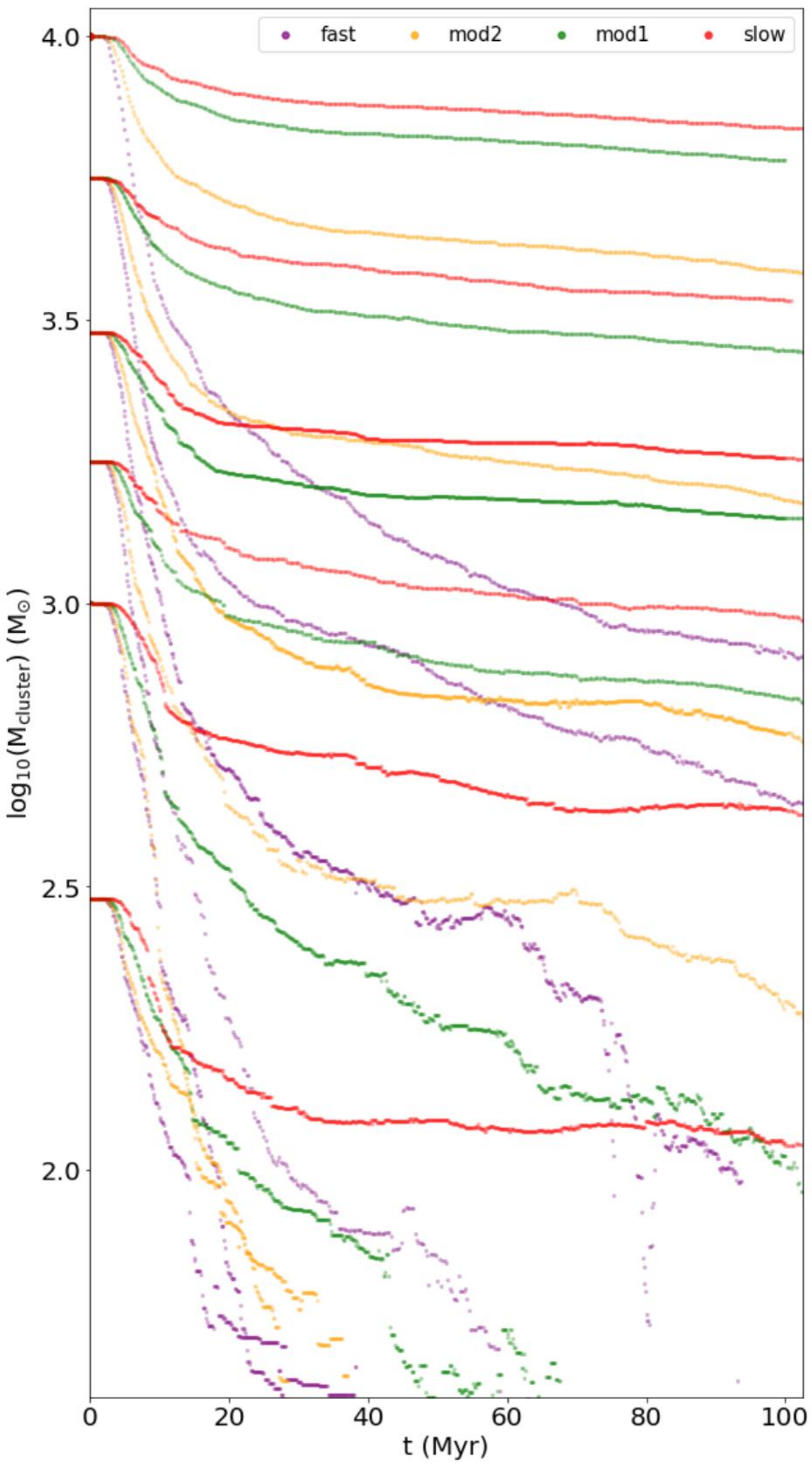}
\caption{Evolution of the cluster mass over time under different gas expulsion modes in the individual cluster simulations. From bottom to top, the initial cluster masses are 300, 1000, 1778, 3000, 5623 and 10000 M$_{\odot}$, respectively.
The timestep is linearly spaced at 0.25 Myr.
}
\label{Mti} 

\end{figure}
\begin{figure*}
\centering
\includegraphics[width=1\textwidth]{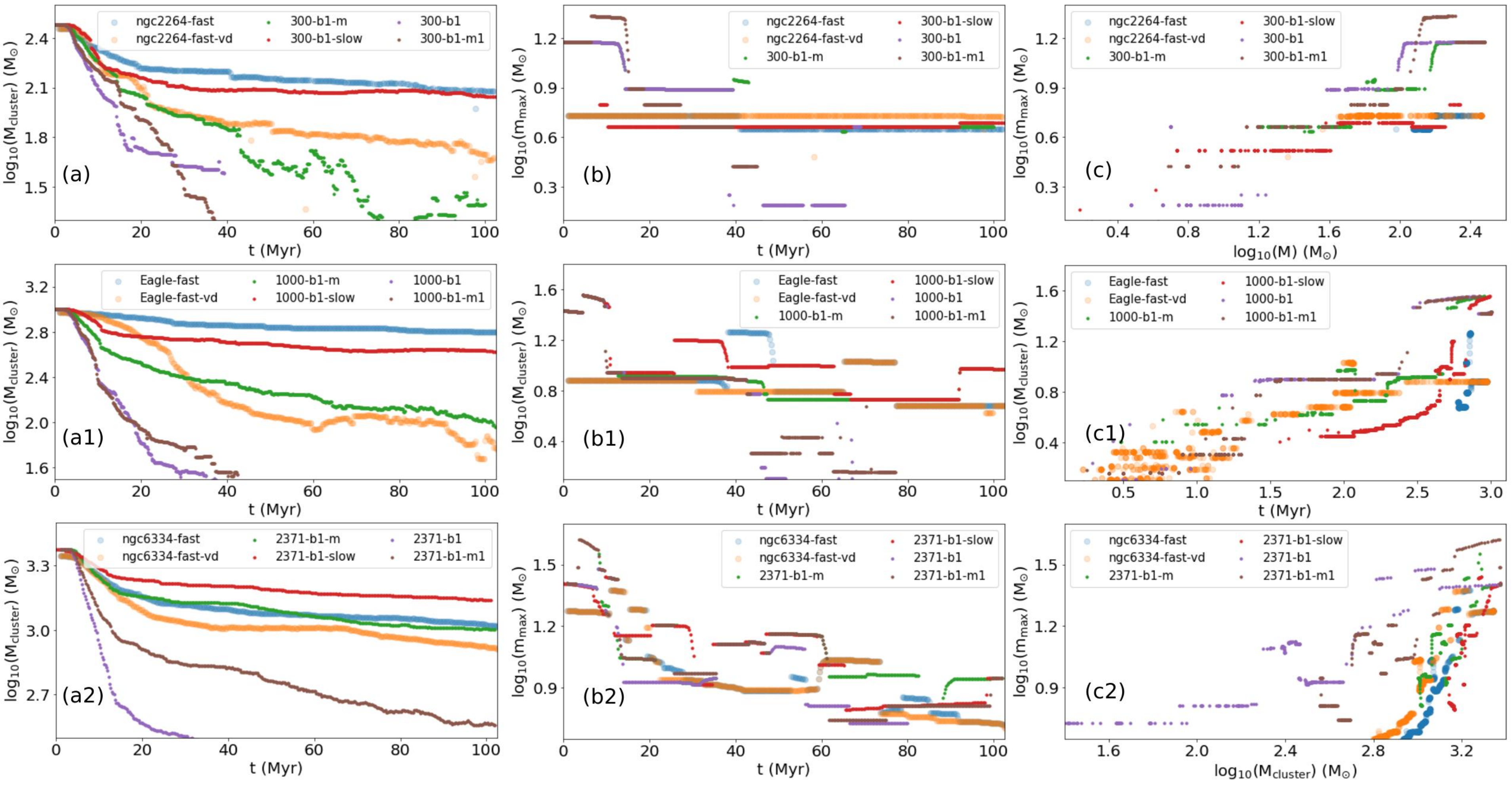}
\caption{Comparison between the evolution of cluster complexes and individual clusters with similar total masses. The masses of cluster complexes "NGC2264", "Eagle", and "NGC6334" are 314, 1015, and 2207 $M_{\odot}$, respectively. From left to right, the panels show the evolutionary tracks in the $M_{\rm cluster}$–t, $m_{\rm max}$–t and $m_{\rm max} -M_{\rm cluster}$ diagrams.
The timestep is linearly spaced at 0.25 Myr.
}
\label{Mtic} 
\end{figure*}

\section{N-body simulations}\label{nbody}

\subsection{Individual cluster simulation}\label{individual}

% We computed three clusters with stellar masses of 300 M$_{\odot}$, 1000 M$_{\odot}$, and 3000 M$_{\odot}$ in the first 100 Myr.
The parameters for the simulation are summarized from previous works (i.e.,
\citealt{Kroupa2001-321,
Baumgardt2007-380,
Banerjee2012-746,
Banerjee2013-764,
Banerjee2014-787,
Banerjee2015-447,
Oh2015-805,
Oh2016-590,
Banerjee2017-597,
Brinkmann2017-600,
Oh2018-481,
Wang2019-484,
Pavlik2019-626,
Dinnbier2022-660,
Zhou2024-691-204,
Zhou2024-691-293}).
The previous simulations have already demonstrated the effectiveness and rationality of the parameter settings (see below). The influence of different parameter settings on simulation results and the discussion of the multidimensional parameter space can also be found in the works cited above. 

The initial density profile of all clusters is the Plummer profile \citep{Aarseth1974-37, HeggieHut2003, Kroupa2008-760}, an appropriate choice since the molecular clouds’ filaments in which stars form have been found to have Plummer-like cross sections \citep{Malinen2012-544,Andre2022-667},
and open star clusters can also be described by the Plummer model \citep{Roeser2011-531,Roeser2019-627}.
Moreover, such a specific initial profile does not significantly affect the overall expansion rate of a cluster, as discussed in \citet{Banerjee2017-597}, 
which is primarily governed by the total stellar mass loss and the dynamical interactions occurring within the inner part of the cluster. 
The half-mass radius, $r_{h}$, of the cluster is given by the $r_{\rm h}-M_{\rm ecl}$ relation \citep{Marks2012-543}:
\begin{equation}
 \frac{r_{\rm h}}{{\rm pc}}=0.10_{-0.04}^{+0.07}\times\left(\frac{M_{\rm ecl}}{M_{\odot}}\right)^{0.13\pm0.04}\;.
 \label{rm}
\end{equation}
All clusters are fully mass segregated ($S$=1), with no fractalization, and in a state of virial equilibrium ($Q$=0.5). $S$ and $Q$ are the degree of mass segregation and the virial ratio of the cluster, respectively. More details can be found in \citet{Kupper2011-417} and the user manual for the \texttt{McLuster} code.
The initial segregated state is detected for young clusters and star-forming clumps/clouds \citep{Littlefair2003-345,Chen2007-134,Portegies2010-48,Kirk2011-727,Getman2014-787,Lane2016-833,
Alfaro2018-478,Plunkett2018-615, Pavlik2019-626, Nony2021-645, Zhang2022-936,
Xu2024-270}, but the degree of mass segregation is not clear. 
In simulations of the very young massive clusters R136 and NGC 3603 with gas expulsion by \citet{Banerjee2013-764}, mass segregation does not influence
the results. In \citet{Zhou2024-691-204}, we compared $S$=1 (fully mass segregated) and $S$=0.5 (partly mass segregated) and found similar results. We also discussed settings with and without fractalization in \citet{Zhou2024-691-204}; the results of the two are also consistent.
The initial mass functions of the clusters are chosen to be canonical \citep{Kroupa2001-322}, with the most massive star following the $m_{\rm max}-M_{\rm ecl}$ relation  of \citet{Weidner2013-434}:
\begin{equation}
\label{eq:mmaxmecl2}
y = a_0 + a_1 x + a_2  x^2 + a_3  x^3 ,
\end{equation}
where $y$ = $\log_{10}(m_\mathrm{max}/M_\odot)$, $x$ = $\log_{10}(M_\mathrm{ecl}/M_\odot)$, $a_0$ = -0.66, $a_1$ = 1.08, $a_2$ = -0.150, and $a_3$ = 0.0084. We assumed the clusters to be at solar metallicity \citep[i.e., $Z=0.02$;][]{von2016ApJ...816...13V}. 
The clusters travel along circular orbits within the Galaxy, positioned at a galactocentric distance of 8.5 kpc, and are moving at a speed of 220 km s$^{-1}$.

The initial binary setup follows the method described in \citet{Wang2019-484}.
All stars are initially in binaries, that is to say, $f_{\rm b}$=1, where $f_{\rm b}$ is the primordial binary fraction.
\citet{Kroupa1995-277-1491,Kroupa1995-277-1507} propose that stars with masses below a few $M_\odot$ are initially formed with a universal binary distribution function and that star clusters start with a $100\%$ binary fraction. Inverse dynamical population synthesis was employed to derive the initial distributions of binary periods and mass ratios. \cite{Belloni2017-471} introduces an updated model of pre-main-sequence eigenevolution, originally developed by \cite{Kroupa1995-277-1507}, to account for the observed correlations between the mass ratio, period, and eccentricity in short-period systems. For low-mass binaries, we adopted the formalism developed in \citet{Kroupa1995-277-1491,Kroupa1995-277-1507} and \citet{Belloni2017-471} to characterize the period, mass ratio, and  eccentricity distributions.
For high-mass binaries (OB stars with masses $>5$~M$_\odot$), we utilized the \citet{Sana2012-337} distribution, which is derived from O stars in OCs. The distribution functions of the period, mass ratio, and eccentricity are presented in \cite{Oh2015-805} and \cite{Belloni2017-471}. 

Accurately modeling gas removal from embedded clusters is challenging due to the complexity of radiation hydrodynamical processes, which involve uncertain and intricate physical mechanisms. 
To simplify the approach, we simulated the key dynamical effects of gas expulsion by applying a diluting, spherically symmetric external gravitational potential to a model cluster, following the method presented in \citet{Kroupa2001-321} and \citet{Banerjee2013-764}. 
This analytical approach is partially validated by \citet{Geyer2001-323}, who conducted comparison simulations using the smoothed particle hydrodynamics method to treat the gas. The hydrodynamics+N-body simulations in \citet{Farias2024-527} also confirm that the exponential decay model presented in equation.\ref{eq:mdecay} generally provides a good description of gas removal driven by radiation and wind feedback.
Specifically, we used the potential
of the spherically symmetric, time-varying mass distribution:
\begin{eqnarray}
M_g(t)=& M_g(0) & t \leq \tau_d,\nonumber\\
M_g(t)=& M_g(0)\exp{\left(-\frac{(t-\tau_d)}{\tau_g}\right)} & t > \tau_d.
\label{eq:mdecay}
\end{eqnarray}
Here, $M_g(t)$ is the total mass of the gas; it is spatially distributed with the Plummer density distribution \citep{Kroupa2008-760} and starts depleting after a delay of $\tau_d$, and is totally depleted 
within a timescale of $\tau_g$. The Plummer radius of the gas distribution is kept time-invariant \citep{Kroupa2001-321}.
This assumption is an approximate model of the effective gas reduction within the cluster in the situation that gas is blown out while new gas is also accreting into the cluster along filaments such that the gas mass ends up being reduced with time but the radius of the gas distribution remains roughly constant. As discussed in \citet{Urquhart2022-510},
the mass and radius distributions of the ATLASGAL clumps at different evolutionary stages are quite comparable.
We used an average velocity of the outflowing gas of $v_g\approx10$ km s$^{-1}$, which is the typical sound speed in an HII region. This gives
$\tau_g=r_h(0)/v_g$,
where $r_h(0)$ is the initial half-mass radius of the cluster. As for the delay time, we take the representative value of $\tau_d\approx0.6$ Myr
\citep{Kroupa2001-321}, this being about the lifetime of the ultracompact HII region. 
As shown in \citet{Banerjee2013-764}, varying the delay time, $\tau_d$, primarily results in a temporal shift in the cluster’s rapid expansion phase, without significantly impacting its subsequent evolution for times greater than $\tau_d$. 
Protoclusters typically form in hub-filament systems \citep{Motte2018-56,Vazquez2019-490, Kumar2020-642,Zhou2022-514}, which are located in hub regions. Compared to the surrounding filamentary gas structures, the hub region, as the center of gravitational collapse, is usually more regular, as shown in \citet{Zhou2022-514,Zhou2024-686-146}. 
Thus, modeling a spherically symmetric mass distribution is appropriate.
% (see Sect. \ref{sm} for more discussion). 

In this work, we assumed a SFE $\approx$ 0.33 as a benchmark (i.e., $M_{g}(0)$ = 2$M_{\rm ecl}(0))$. This value has been widely used in the simulations cited above and has been proven effective in reproducing the observational properties of stellar clusters. Such a SFE is also consistent with the value obtained from hydrodynamical calculations that include self-regulation \citep{Machida2012-421,Bate2014-437} and as well with observations of embedded systems in the solar neighborhood \citep{Lada2003-41,Megeath2016-151}.
In \citet{Zhou2024-688L}, by comparing the mass functions of the ATLASGAL clumps and the identified embedded clusters, we found that a SFE of $\approx$ 0.33 is typical for the ATLASGAL clumps. 
In \citet{Zhou2024PASP-1}, assuming SFE = 0.33, it was shown that the total bolometric luminosity of synthetic embedded clusters can fit the observed bolometric luminosity of ATLASGAL clumps with HII regions. In \citet{Zhou2024PASP-2}, we directly calculated the SFE of ATLASGAL clumps with HII regions and found a median value of $\approx$0.3.

% \subsubsection{Slow and moderate gas expulsions}\label{sm}

Embedded clusters form in clumps. More massive clumps can produce more massive clusters, leading to stronger feedback and a higher gas expulsion velocity \citep{Dib2013-436}. 
There should be correlations between the feedback strength, the clump (or cluster) mass, and the gas expulsion velocity ($v_g$). And low-mass clusters should have a slower gas expulsion process compared with high-mass clusters. As shown in \citet{Pang2021-912}, low-mass clusters tend to agree with the simulations of slow gas expulsion. Except for the feedback strength, the SFE determines the total amount of the remaining gas, which also influences the timescale of gas expulsion. The gas expulsion process is driven by feedback, and the effectiveness of the feedback will depend on the geometric shape of the gas shell surrounding the embedded cluster  \citep{Wunsch2010-407,Rahner2017-470}. Therefore, a complex or nonspherical gas distribution would also change the timescale of gas expulsion. Moreover,
the total amount of the residual gas not only affects the gas expulsion timescale, but also significantly influences the strength of the gas potential. The strength of the external gas potential may have a considerable impact on the evolution of embedded star clusters. The total amount of residual gas is determined by the SFE of the clump. As verified in \citet{Zhou2024-691-293}, a lower SFE is equivalent to a shorter gas expulsion timescale. 
In short, the uncertainty of the above parameters can ultimately be incorporated into the timescale of gas expulsion. Thus, apart from the fast gas expulsion with the gas decay time, $\tau_g$, described above, we also simulated slow and moderate gas expulsions. 
For the slow gas expulsion, the gas decay time was set to 10$\tau_g$ \citep{Zhou2024-691-293}. 
The moderate gas expulsion is between the fast and slow gas expulsions. Considering the large parameter space between $\tau_g$ and 10$\tau_g$, we simulated two kinds of moderate gas expulsions: 2$\tau_g$ ("moderate2") and 5$\tau_g$ ("moderate1"). 

\subsection{Procedure}

The \texttt{McLuster} program \citep{Kupper2011-417} was used to set the initial configurations. 
The dynamical evolution of the model clusters was computed using the state-of-the-art \texttt{PeTar} code \citep{Wang2020-497}. 
\texttt{PeTar} employs well-tested analytical single and binary stellar evolution recipes (SSE/BSE)
\citep{Hurley2000-315,Hurley2002-329,Giacobbo2018-474,Tanikawa2020-495,Banerjee2020-639}. 

\subsection{Cluster coalescence simulation}

% For the 17 MSFRs investigated in the MYStIX project, nine of them are subcluster complexes. As shown in \citet{Kuhn2014-787}, some of these nine MSFRs exhibit similar morphologies. Finally, we selected NGC 1893, NGC 6334 and the Carina nebula as representatives. 
% In these three MSFRs, the spatial distribution of their subclusters varies, particularly with significant differences in the total mass of the subclusters, which helps to fit the observed physical parameters of open clusters.
For three massive star-forming regions (NGC 1893, NGC 6334 and the Carina nebula),
we have identified their subclusters using the dendrogram algorithm according to the surface density distributions of stars in \citet{Zhou2024-688L}, and calculated their physical parameters. Considering that
the mass-radius relation of embedded clusters can be well fitted by the $\approx$1 Myr expanding line in \citet{Zhou2024-691-204}, 
we simulated the evolution of the subclusters in each MSFR for the first 1 Myr using the recipe described in Sec.\ref{individual}. The initial masses of the subclusters strictly follow the observed values, and each subcluster is simulated individually.
Then, all subclusters were collected together to create an initial configuration similar to the observation.
% , as shown in Fig.\ref{cluster-merger-example}. 
% The separations between subclusters in Fig.\ref{cluster-merger-example}(b) strictly adhere to the observed values. 

We first assume all embedded clusters are initially at rest and located on the same plane ("fast"). A more realistic scenario is that clusters have relative velocities and line-of-sight spatial separations ("fast-vd"). To take these two factors into account, we approximate the molecular cloud containing the clusters as an ellipsoid. More details can refer to \citet{Zhou2025-537-845}.
% To derive the lengths of the three axes of the ellipsoid (semi-major axis, $a$, intermediate axis, $b$, semi-minor axis, $c$), we fit the coordinates of the clusters located at the periphery to determine $a$ and $b$. 
% Given that the real cloud shape should be more sheet-like instead of spherical \citep{Shetty2006-647,Inutsuka2015-580,Arzoumanian2018-70,Kohno2021-73,Arzoumanian2022-660,Rezaei2022-930,Zhou2023-519,Zhou2023-676,Clarke2023-519,Ganguly2023-525}, we assume $c=b$.
Then we randomly distribute clusters along the Z-axis (line-of-sight). 
% As shown in Fig.\ref{cluster-merger-example}, Fig.\ref{cluster-merger-example1} and Fig.\ref{cluster-merger-example2},
% the spatial distribution of clusters is nearly spherical. Therefore, we have overestimated the spacing between clusters.
% For molecular clouds in the catalog of \citet{Miville2017-834} across the Galactic plane, the most probable cloud size is $\approx$30 pc, and the most probable value of the velocity dispersion is $\approx$1.95 km s$^{-1}$ on the scale of clouds. 
% In our case, the sizes of the MSFRs are $\approx$5 pc for NGC 6334 and NGC 1893, and $\approx$20 pc for the Carina MSFR.
% As a conservative estimate
As discussed in \citet{Zhou2025-537-845}, we take the velocity dispersion of the original molecular clouds in the three MSFRs as 2 km s$^{-1}$ and assume that the clusters inherit the velocity dispersion of the clouds. The system's center has a velocity of 0 km s$^{-1}$, and the velocity of the outermost cluster is 2 km s$^{-1}$. The velocities of other clusters are distributed according to the Larson relation, i.e. $v \propto L^{0.5}$, where $L$ is the distance to the system's center. 

\end{appendix}

\end{document}